%% file: cnl.tex
\documentclass[reqno]{amsart}
\usepackage{comment,amssymb,latexsym,cite,upref}
\usepackage{finalT}

\numberwithin{equation}{section}

\input cnl.def

\begin{document}

\input cnl_tit.tex
\input intro.tex
\input ee.tex
\input idata.tex

\input limit.tex
\input nlim.tex

\bigskip



\input cnl.bbl

\end{document}

%% file: cnl_tit.tex
\title[Cosmological Newtonian limit without averaging]{A rigorous formulation of the cosmological Newtonian limit without averaging}
\author[T.A. Oliynyk]{Todd A. Oliynyk}
\address{School of Mathematical Sciences\\
Monash University, VIC 3800\\
Australia}
\email{todd.oliynyk@sci.monash.edu.au}

\begin{abstract}
\noindent We prove the existence of a large class of one-parameter families of cosmological solutions to the Einstein-Euler
equations that have a Newtonian limit. This class includes solutions that represent a finite, but otherwise arbitrary,
number of compact fluid bodies. These solutions provide exact cosmological models that admit Newtonian limits but, are
not, either implicitly or explicitly, averaged.
\end{abstract}

\maketitle 

%% file: intro.tex
\sect{intro}{Introduction}
Gravitating perfect fluids are governed by the Einstein-Euler equations
\eqn{EEeqnA}{
G^{ij} = \frac{8\pi G}{c^4} T^{ij}\,
\AND \nabla_{i} T^{ij} = 0,} where \eqn{EEdefsA}{ T^{ij} = (\rho + c^{-2}p)v^i v^j + pg^{ij},  } with $\rho$ the fluid density,
$p$ the fluid pressure, $v^i$ the fluid
four-velocity normalized
by $v^i v_i = -c^2$, $c$ the speed of light, and $G$ the
Newtonian gravitational constant.
By suitably rescaling (see \cite{Oli06}), these
equations can be written as
\leqn{EEeqn}{ G^{ij} = 2\ep^4 T^{ij} \AND
\nabla_{i} T^{ij} = 0, } where \eqn{EEdefs}{ T^{ij} = (\rho+\ep^2 p)v^i v^j + p g^{ij} \AND v^i v_i = -\frac{1}{\ep^2}. }
In the article \cite{Oli09}, we proved the existence of a large class of one-parameter families of solutions to
this system defined for $0 < \ep <\ep_0$ that \
\begin{itemize}
\item[(i)] exist on a common piece of spacetime of the form $M = [0,T)\times \Tbb^3$,
\item[(ii)] converge as $\ep \searrow 0$ to solutions of the cosmological Poisson-Euler equations of Newtonian gravity, and
\item[(iii)] are differentiable in $\ep$ to any prescribed order $\ell\in \Nbb$.
\end{itemize}
The properties (i)-(iii) guarantee that these one parameter families of solutions to the Einstein-Euler equations have  valid Newtonian limits
and admit post-Newtonian expansions to order $\ell/2$. However, in order to establish the existence of these solutions, we required
that the two following conditions are satisfied by the initial fluid density $\rho|_{t=0} = \rho_0$ and fluid three-velocity
$v^I|_{t=0} = w^I_0$:
\leqn{twocond1}{
\rho_0(\xv) > 0 \quad \forall \; \xv\in \Tbb^3,
}
and
\leqn{twocond2}{
\int_{\Tbb^3} \rho_0 w^I_0 \, d^3 x = 0.
}
Here, we are using  $\xv = (x^I)$ $(I=1,2,3)$ to denote the standard periodic
coordinates with period $1$ on the torus $\Tbb^3= S^1\times S^1 \times S^1$, and $t=x^0$ to denote an absolute Newtonian time coordinate
on the interval $[0,T)$.

The main aim of this article is to remove the conditions \eqref{twocond1}-\eqref{twocond2} on the initial data. Although, we
do not discuss the existence of post-Newtonian expansions in this article, it is not difficult to see that the results of this article
can be combined with those of \cite{Oli09} to prove the existence of post-Newtonian expansions to arbitrary order that do not satisfy
\eqref{twocond1}-\eqref{twocond2}.

From our point of view, there are two important reasons for removing the conditions \eqref{twocond1}-\eqref{twocond2}. This first
reason is that \eqref{twocond2} is an exact averaging condition that cannot be expected to be exactly satisfied for real
systems. The second is that by removing \eqref{twocond1}, we  can choose an initial fluid density of the form
\eqn{iden1}{
\rho_0 = \sum_{\nu=0}^N \rho_{0,\nu}
}
where $N$ is any integer, and
\eqn{iden2}{
\text{dist}\bigl(\text{supp} \, \rho_{0,\nu}, \text{supp}\, \rho_{0,\mu} \bigr) > 0  \quad \nu \neq \mu.
}
This initial data represents a finite but otherwise arbitrary number of compact fluid bodies (i.e. stars).
Thus, solutions generated by this initial data can be used to model an arbitrary collection of stars, and therefore provide
an exact model for the universe. Consequently, we obtain one-parameter families of cosmological solutions that admit
Newtonian limits, and are not,
either implicitly or explicitly, averaged.

To formulate the Newtonian limit, we require (see \cite{Oli09}) the
following FLRW dust solution $(g^\ep_{ij}=h^\ep_{ij},v^i_\ep=\xi^i,\rho_\ep = \mu_\ep,p_\ep=0)$
of \eqref{EEeqn}:
\lalign{flrw1}{
h^\ep_{ij} & = -\frac{1}{\ep^2}\delta_i^0\delta_j^0
+ a_\ep\delta_i^I\delta_j^J\delta_{IJ} , \label{flrw1.1} \\
\xi^i & = \delta^i_0, \label{flrw1.2} \\
\mu_\ep & = \frac{3}{8} \left[\frac{a_\ep'}{a_\ep}\right]^2, \label{flrw1.3}
}
where $a_\ep = a_\ep(t)$ and $\mu_\ep = \mu_\ep(t)$ satisfy the differential equations
\lalign{flrw2}{
&\frac{a_\ep''}{a_\ep} -\frac{1}{2}\left[\frac{a_\ep'}{a_\ep}\right]^2 + \frac{2}{3}\mu_\ep = 0,
\label{flrw2.1}
\intertext{and}
& \mu_\ep' + \frac{3}{2} \frac{a_\ep'}{a_\ep} \mu_\ep = 0 \label{flrw2.2},
}
respectively. Here, we are using
\eqn{dt}{
(\cdot ) ' = \frac{d\;}{dt}(\cdot).
}
The differential equations \eqref{flrw2.1}-\eqref{flrw2.2} can be integrated explicitly to give
\lalign{flrw3}{
a_\ep(t) &= a_{\ep}(0)\left( \sqrt{\frac{3\sigma(\ep)}{2}}t+ 1 \right)^{\frac{4}{3}}, \label{flrw3.1}
\intertext{and}
\mu_\ep(t) &= \frac{2}{3}\left(t+\sqrt{\frac{2}{3\sigma(\ep)}}\right)^{-2}, \label{flrw3.2}
}
where $a_\ep(0)$ and $\sigma(\ep)=\mu_\ep(0)$ are arbitrary (positive) functions that are analytic in a neighborhood of $\ep=0$. We fix the arbitrary length scale by setting $a_\ep(0) = 1$, while $\sigma(\ep)$
will be determined by later considerations. Also, to reduce notation we will drop the $\ep$ from the function $a$ and $\mu$ except in situations where we want to emphasize the $\ep$-dependence.

In this article, we use a slight variation of the approach used in \cite{Oli09} to analyze the limit $\ep \searrow 0$. The first
step in the analysis is to replace
the metric $g_{ij}$ and the fluid velocity
$v^i$ with variables that are compatible with the limit
$\ep \searrow 0$. The new gravitational variable $\ub^{ij}$ is defined by
\leqn{metrecA}{
g^{ij} = \frac{q^{ij}}{\sqrt{-|h|\det{q^{kl}}}} \qquad (|h| = -\det (h_{ij}) )
}
where
\leqn{metrecB}{
q^{ij} = h^{ij} + \ep^2 J^i_k J^j_l \ub^{kl} \AND J^j_i = \ep \delta^j_0\delta^0_i + \delta^j_I \delta^I_i,
}
while the new fluid four-velocity $w^i$ is defined by
\leqn{wdef.intro}{
v^i = (1 + \ep w^0) \delta^i_0 + \delta^i_I w^I.
}

As in \cite{Oli06,Oli07}, we use Makino's technique \cite{Mak,Ren92} to generate
perfect fluid solutions with compact support. This requires
the use of an isentropic equation of
state of the form
 \leqn{eos}{ p =
K\rho^{(n+1)/n}, } where $K \in \Rbb_{>0}$, $n\in \Nbb$. This allows
for the introduction of the density variable
\leqn{dendef}{ \rho = \frac{1}{\bigl(4Kn(n+1)\bigr)^n}\alpha^{2n} }
which is used to formulate the Euler equations as
a symmetric hyperbolic system that
is regular across the
fluid-vacuum interface.
In this way,
it is possible to construct solutions to the Einstein-Euler
equations that represent compact gravitating fluid bodies (i.e. stars)
both in the Newtonian and
relativistic setting \cite{Mak,Ren92}.
Although, these solutions do
not include static stars of finite radius \cite{Ren92},  they are general enough
to understand the mathematical issues involved in the Newtonian
limit.

The main result of this article is to show that solutions to the  following equations, which we  refer to
as the \emph{cosmological Poisson-Euler-Makino equations}, rigorously approximate fully relativistic solutions to the Einstein-Euler
equations \eqref{EEeqn} up to an error term of order $\ep$ as measured in suitable Sobolev spaces:
\lalign{limA}{
\del_t \alphat & = -\wt^I\del_I \alphat - \frac{\alphat}{2n} \del_I \wt^I - \frac{3\at'}{4n\at} \alphat, \label{limA.1} \\
\del_t \wt^J & = -\wt^I\del_I \wt^J - \frac{\alphat}{2n\at}\del^J\alphat -\frac{\at'}{\at}\wt^J + g^J, \label{limA.2} \\
\Delta \Phit & = 4 \at (\rhot - \mut) \qquad \bigl(\rhot = \bigl(4Kn(n+1)\bigr)^{-n}\alphat^{2n}\bigr), \label{limA.3}
}
where
\eqn{muatdef}{
\at = a_0, \qquad \mut = \mu_0,
}
\leqn{gacc1}{
g^J = -\frac{1}{\at}\left(\frac{3}{2}\frac{\at'}{\at} \frac{1}{\mut} \int_{\Tbb^3} \rhot\wt^J \, d^3 x
+ \frac{1}{4}\del^J\Phit \right),
}
and $\ip{\cdot}{\cdot}_{L^2}$ is the standard $L^2$ inner-product on $\Tbb^3$, i.e.
\eqn{L2}{
\ip{\psi_1}{\psi_2}_{L^2} = \int_{[0,1]^3} \psi_1(\xv)\psi_2(\xv)\, d^3 x.
}
Here, and for the rest of the article,
\eqn{flatLap}{\Delta = \delta^{IJ}\del_I\del_J}
will denote the flat
Laplacian, and we set
\eqn{delraise}{
\del^J = \delta^{IJ}\del_I.
}

In terms of the fluid density $\rhot$, equations
\eqref{limA.1}-\eqref{limA.3} take the form
\lalign{limB}{
\del_t \rhot & = -\wt^I\del_I \rhot - \rhot\del_I \wt^I - \frac{3}{2}\frac{\at'}{\at}\rhot , \label{limB.1} \\
\del_t \wt^J & = -\wt^I\del_I \wt^J - \frac{1}{\at\rhot}\del^J p(\rhot) -\frac{\at'}{\at}\wt^J + g^J, \label{limB.2} \\
\Delta \Phit & = 4 \at (\rhot - \mut), \label{limB.3}
}
which we refer to as the
\emph{cosmological Poisson-Euler equations}. We note that these equations agree with the Newton-Cartan field equations
for a gravitating fluid formulated in \emph{adapted coordinates} \cite{Kunz76,RS97}.

An important point is that the constant $\sigma(0) = \mut(0)$ is not arbitrary but is determined by
the initial density
\eqn{rhoidata}{
 \rhot_0 = \rhot|_{t=0}
}
according to
\leqn{mufix1}{
\sigma_0 := \mut(0) = \int_{\Tbb^3}\rhot_0\, d^3 x > 0.
}
From equations  \eqref{flrw2.2} and \eqref{limB.1}, it is not difficult to see that this average is preserved
under evolution
\leqn{mufix6}{
\mut(t) = \int_{\Tbb^3} \rhot(t)\,  d^3 x,
}
which in turn implies (see \eqref{flrw1.3} ) that
\leqn{mufix7}{
\at(t) = \exp\left( \int_{0}^{t}\left(\frac{8}{3} \int_{\Tbb^3} \rhot(s)\,  d^3 x\right)^{\frac{1}{2}} ds \right).
}
Also, setting
\leqn{zdef}{
\zeta^J = \int_{\Tbb^3} \rhot \wt^J \, d^3 x
}
it follows directly from the cosmological Poisson-Euler equations \eqref{limB.1}-\eqref{limB.3} that $\zeta^J$
satisfies the equation
\eqn{zeqn1}{
{\zeta^J}' = -4\frac{\at'}{\at} \zeta^J
}
which can be integrated to give
\leqn{zeqn2}{
 \zeta^J(t) = \frac{1}{\at(t)^4} \int_{\Tbb^3}\rhot_0 \wt^J_0 \, d^3 x,
}
where
\eqn{wJidata}{
\wt_0^J = \wt^J|_{t=0}
}
is the initial fluid 3-velocity. Using \eqref{zeqn2}, the acceleration due to gravity
\eqref{gacc1} can be reexpressed in the form
\eqn{gacc2}{
g^J = -\frac{1}{\at}\left(\frac{4}{\at'\at^3} \int_{\Tbb^3}\rhot_0 \wt^J_0 \, d^3 x
+ \frac{1}{4}\del^J\Phit \right).
}

For purposes of interpretation, it is often useful to introduce \emph{Galilei coordinates} \cite{Kunz76,RS97}.
These coordinates are defined as follows:
suppose $\{\rhot(t,x),w^I(t,x),\Phi(t,x)\}$ is a solution of the cosmological
Poisson-Euler equations \eqref{limB.1}-\eqref{limB.3} on $M = [0,T)\times \Tbb^3$. Then, letting
$\Mt = [0,T)\times \Rbb^3$ denote the covering space, we define a diffeomorphism on $\Mt$ by
\eqn{phihdef}{
\psi \: : \: \Mt \longrightarrow \Mt \; : \: (t,x) \longmapsto (t,x/\sqrt{\at(t)}).
}
Lifting the cosmological Poisson-Euler equations to $\Mt$, and then pulling back by $\psi$ shows that\footnote{In the Newton-Cartan theory, the
fluid velocity 3-vector $\wt^I$ is the spatial part of a 4-vector $\wt = \del_t + \wt^I\del_I$ \cite{RS97}. The formula
\eqref{hatvars1.2} follows from the calculating the spatial components of $\wh = \psi^* \wt$. The other two formulas
\eqref{hatvars1.1} and \eqref{hatvars1.3} follow from the definition of the pullback, i.e.
$\rhoh = \psi^*\rhot$ and $\Phih = \psi^*\Phit$.}
\lalign{hatvars1}{
\rhoh(t,x) &= \rhot\bigl(t,x/\sqrt{\at(t)}\bigr), \label{hatvars1.1} \\
\wh^J(t,x) & = \sqrt{\at(t)}\wh^J\bigl(t,x/\sqrt{\at(t)}\bigr)+ \frac{1}{2}\frac{\at'(t)}{\at} x^J, \label{hatvars1.2}\\
\Phih(t,x) &= \Phit\bigl(t,x/\sqrt{\at(t)}\bigr), \label{hatvars1.3}
}
satisfy
\lalign{limC}{
\del_t \rhoh & = -\wh^I\del_I \rhoh - \rhoh\del_I \wh^I, \label{limC.1} \\
\del_t \wh^J & = -\wh^I\del_I \wh^J - \frac{1}{\rhoh}\del^J p(\rhoh) + \gh^J, \label{limC.2} \\
\Delta \Phih & = 4(\rhoh - \mut), \label{limC.3}
}
where
\leqn{hatvars2}{
\gh^J = - \frac{4}{\at'\at^{7/2}} \int_{\Tbb^3}\rhot_0 \wt^J_0 \, d^3 x
- \frac{1}{4}\del^J\Phih -\frac{\mut}{3} x^J .
}
A Newtonian potential can be defined by
\leqn{hatvars3}{
\check{\Phi} = \frac{\Phih}{4} + \frac{\mut}{6}\delta_{IJ}x^I x^J +
\frac{4}{\at'\at^{7/2}}\delta_{IJ}x^I\int_{\Tbb^3}\rhot_0 \wt^J_0 \, d^3 x.
}
This potential satisfies the Poisson equation
\leqn{hatvars4}{
\Delta \check{\Phi} = \rhoh,
}
while the acceleration due to gravity $\gh^J$ takes the familiar
form
\leqn{hatvars5}{
\gh^J = - \del^J \check{\Phi}.
}
Together, equations \eqref{limC.1}, \eqref{limC.2}, \eqref{hatvars4}, and \eqref{hatvars5} show that
solutions to the cosmological Poisson-Euler equations determine solutions to the standard Poisson-Euler
equations on the covering space $\Mt$.

\subsect{not}{Notation} Before proceeding, we first fix our notation and introduce a number of
function spaces that will be used in this article.
Given a finite dimensional vector space $V$, we let $H^s(V)$ denote the standard  Sobolev space of $V$-valued
maps on $\Tbb^3$. When $V=\Rbb$, we just write $H^s$. The only two vector spaces that will be used in
this article are  $\Rbb^N$ and the space of symmetric matrices $\Sbb{N} = \{ \: (u^{ij})\in \Mbb{N}\: |\: u^{ij} = u^{ji} \: \}$.

We denote the projection operator
onto the $L^2$ orthogonal complement of the constant function $1$ by
\eqn{L2proj}{
\Pi(\psi) = \psi - \ip{1}{\psi}_{L^2}1 \quad \forall \; \psi\in L^2(\Tbb^3).
}
Given $\{\ev_{\alpha}\}_{\alpha=1}^N$ any basis for $V$, we use this projection to define
\eqn{Hsbar}{
\Ho^s(V)= \Bigl\{\, \psi = \sum_{\alpha=1}^N \psi^\alpha \ev_\alpha \in H^s(V) \: \Bigl|\:  \ip{1}{\psi^\alpha}_{L^2} = 0  \Bigr\}.
}
We also define the standard hyperbolic evolution spaces
\eqn{Xspace}{
X_{T,\ell,s}(V) = \bigcap_{p=0}^{\ell+1} C^p\bigl([0,T), H^{s-p}(V)\bigr).
}
and write $X_{T,\ell,s}$ if $V=\Rbb$.

\subsect{main}{Main result}

The main results of this article is the following Theorem which establishes the existence
of a wide class of 1-parameter families of solutions to the Einstein-Euler equations that
converge in the limit $\ep \searrow 0$ to solutions of the cosmological Poisson-Euler
equations \eqref{limB.1}-\eqref{limB.3}. See Section \ref{nlim} for the proof.

\begin{thm} \label{nlimA} \mnote{[nlimA]}
Suppose $K>0$, $n,s,\ell \in \Zbb_{\geq 0}$, $s\geq 3+\ell$, $\alphat_0\in H^s$, $\wt^I_0\in H^s(\Rbb^3)$, $\uc^{IJ}_0 \in \Hb^{s}(\Sbb{3})$ ,
$\uc^{IJ} \in \Hb^{s+1}(\Sbb{3})$, $T_0$ as defined in Proposition \ref{limA}, and
$\sigma_0>0$ is defined by \eqref{mufix1} where $\rhot_0 = (4Kn(n+1))^{-n}\alphat_0^{2n}$. Then
for $\ep_0$ small enough, there exists a $T\in (0,T_0)$ independent of $\ep \in (0,\ep_0)$,
and maps
\gath{nlimA1}{
\ub^{ij}_\ep \in X_{T,\ell+1,s+1}(\Sbb{4}) \qquad 0<\ep<\ep_0,\\
\alpha_\ep \in X_{T,\ell,s}, \quad w^i_\ep(t) \in  X_{T,\ell,s}(\Rbb^4) \qquad 0<\ep<\ep_0, \\
\alphat \in X_{T_0,\ell,s}, \quad \wt^I \in X_{T_0,\ell,s}(\Rbb^3), \quad \Phit \in X_{T_0,\ell+2,s+2}, \\
\sigma \in C^\omega((-\ep_0,\ep_0),\Rbb_{>\sigma_0/2}), \quad \sigma(0) = \sigma_0, \\
y^J \in C^\omega((-\ep_0,\ep_0),\Rbb^3), \quad y^J(0) = -\frac{1}{\sigma_0}\int_{\Tbb^3}\rhot_0 \wt^J_0\, d^3 x
}
such that
\begin{itemize}
\item[(i)]
\alin{nlimA2}{
\alpha_\ep|_{t=0} & = \alphat_0, \\
(w^j_\ep)|_{t=0} &= (w^0_\ep,\wt^I_0), \\
(\ub^{ij}_\ep)\bigl|_{t=0} & = \begin{pmatrix} \uc^{00}_\ep & \frac{1}{\ep} y^J(\ep) + \ub^{0J}_\ep \\
\frac{1}{\ep}y^J(\ep) + \uc^{J0}_\ep & \ep \uc^{IJ} \end{pmatrix},\\
(\del_t\ub^{ij}_\ep)\bigl|_{t=0} &= \begin{pmatrix} \del_t\ub^{00}_\ep\bigl|_{t=0}  & \del_t\ub^{0J}_\ep|_{t=0} \\
\del_t\ub^{0J}_\ep\bigl|_{t=0} & \uc_0^{IJ} \end{pmatrix}
}
where
\alin{nlimA3a}{
\del_t\ub^{00}_\ep\bigl|_{t=0} = -\frac{1}{\ep}\del_I \uc^{I0}_\ep - \frac{3}{2}\ln(a)' \uc^{00}_\ep -\frac{1}{2} a' \delta_{IJ} \ep \uc^{IJ},\\
\del_t\ub^{0J}_\ep\bigl|_{t=0} = -\del_I\uc^{IJ} - \frac{5}{2}\ln(a)'
\left(\frac{1}{\ep}y^{J}(\ep) + \uc^{0J}_\ep\right),
}
\item[(ii)]
$\{\ub_\ep^{ij}(t,\xv),\alpha_\ep(t,\xv),w^i_\ep(t,\xv) \}$
determines, via the formulas  \eqref{metrecA},  \eqref{wdef.intro}, and \eqref{dendef}, a $1$-parameter family of unique solutions
to the Einstein-Euler equations \eqref{EEeqn} in the harmonic gauge
on the common spacetime region $(t,\xv)\in M=[0,T)\times \Tbb^3$,
\item[(iii)]
$\{\Phit(t,\xv),\rhot(t,\xv),\wt^I(t,\xv)\}$ solves the cosmological Euler-Poisson
equations \eqref{limB.1}-\eqref{limB.3}
on the spacetime region $M$, and
\item[(iv)]
there exists a constant $C>0$ independent of $\ep \in (0,\ep_0)$
such that
\alin{newtA4a}{
\ep^{-1}\norm{v^0_\ep(t)-1}_{H^{s-1}}+ &\norm{v^I_\ep(t)-\wt^I(t)}_{H^{s-1}}+ \norm{\rho_\ep(t) -
\rhot(t)}_{H^{s-1}} \leq
C\ep,}
\eqn{newtA4b}{
\norm{\ep\ub_\ep^{ij}(t)-\ut^{ij}(t)}_{H^{s-1}} + \norm{\del_K\ub_\ep^{ij}(t)-\delta^i_0\delta^j_0 \Phit(t)}_{H^{s-1}} \leq C\ep,
}
and
\alin{newtA4c}{
 \norm{\ep \del_t \ub^{00}_\ep(t)}_{H^{s-1}} &\leq C \ep, \\
 \norm{\ep \del_t \ub^{0J}_\ep(t) -  \ut^{0J}_{0}(t)  + \frac{a'(t)}{2a(t)} \ut^{0J}(t)}_{H^{s-1}} & \leq C\ep, \\
\norm{\ep \del_t \ub^{IJ}_{\ep}(t) - \ut^{IJ}_0 (t)}_{H^{s-1}} &\leq C\ep,
}
for all $(t,\ep)\in [0,T)\times (0,\ep_0)$, where
\alin{newtA4c}{
\ut^{ij}_0 & =2\delta_0^{(i}\delta_J^{j)} \left( -\left(\frac{1}{\mut}\zeta^J\right)' - \frac{\at'}{2 \at}\frac{1}{\mut}\zeta^J\right), \\
\ut^{ij} & = 2\delta_0^{(i}\delta_J^{j)}\left( -\frac{2}{\mut}\zeta^J\right),
}
and  $\mut$, $\at$, and $\zeta^J$ are defined by equations \eqref{mufix6}, \eqref{mufix7}, and \eqref{zdef}.
\end{itemize}
\end{thm}

%% file: ee.tex
\sect{edeqns}{The Einstein-Euler equations}

In this section, we adapt the formulation used in \cite{Oli09}
to write the Einstein-Euler equations in a form suitable
to analyze the limit $\ep \searrow 0$ for initial data that
does not satisfy \eqref{twocond1}-\eqref{twocond2}. The main change from
the formalism used in \cite{Oli09} is that scale factor $a_\ep(t)$ appearing
in the FLRW metric is now $\ep$-dependent through the function $\sigma(\ep)$.
This additional freedom allows us to construct initial data that is more general
than that in \cite{Oli09}.

\subsect{red}{Reduced Einstein Equations}

To derive a suitable symmetric hyperbolic system for
the gravitational field equations, we introduce new
coordinates related to old ones by the rescaling
\eqn{bcoords}{ \xb^0 = x^0/\epsilon, \quad \xb^J = x^J,} and let
\eqn{bpartial}{
\quad \delb_i =  \frac{\partial\;}{\partial \xb^i} \, . }
In
the new coordinates, the spacetime metric $\gb_{ij}$ and the FLRW metric $\hb_{ij}$ (see \eqref{flrw1.1})
are given by \eqn{bmetricdef}{ \gb_{ij} = J^k_i J^k_j g_{ij} \AND
\hb_{ij} = -\delta^0_i\delta^0_j + a\delta^I_i\delta^J_i \delta_{IJ}
}
respectively, where $J^j_i$ is defined in \eqref{metrecB}.
The non-zero independent components of the Christoffel symbols $\gammab^k_{ij}$ and the curvature
$\Rcb_{ijkl}$ of the metric $\hb_{ij}$ are:
\lalign{hcurv}{
\gammab^I_{0I} &= \frac{\ep}{2}\frac{a'}{ a}, \label{hcurv.1} \\
\gammab^0_{II} & = \frac{\ep}{2} a', \label{hcurv.2} \\
\Rcb_{0I0I} &= -\frac{\ep^2}{4}\frac{2 a a'' - (a')^2}{a}, \label{hcurv.3}
\intertext{and}
\Rcb_{2121} &=\Rcb_{1313}=\Rcb_{2323} = \frac{\ep^2}{4}(a')^2. \label{hcurv.4}
}

As discussed in the introduction, we take the symmetric 2-tensor $\ub^{ij}$ as our primary gravitational variable
where
\leqn{ubdef}{
\ub^{ij} =\frac{1}{\ep^2} \left( \frac{\sqrt{|\gb|}}{\sqrt{|\hb|}}\gb^{ij}-\hb^{ij}\right)
}
and
\leqn{voldef}{
|\gb| = -\det(\gb_{ij}) \AND |\hb| = -\det(\hb_{ij}) = a^3.
}
The metric can be recovered from the $\ub^{ij}$ by the formula
\leqn{utog}{
\gb^{ij} = \frac{1}{\sqrt{-|\hb|\det(\gt^{kl})}}\gt^{ij},
}
where
\leqn{gtdef}{
\gt^{ij} = \hb^{ij} + \ep^2 \ub^{ij}.
}

Substituting \eqref{utog} into the standard formula for the Christoffel symbols
gives
\lalign{christZ}{
\Gammab^k_{ij} = \gammab^k_{ij}& +\ep^2\Bigl(-\gt_{l(i}\Db_{j)}\ub^{kl}+\Half\gt^{kl}\gt_{im}\gt_{jn}\Db_l\ub^{mn} -\Quarter
   \gt^{kl}\gt_{ij}\gt_{mn}\Db_l\ub^{mn} +\Half\gt_{lm}\delta^k_{(i}\Db_{j)}\ub^{lm}\Bigr), \label{christ}
}
where $(\gt_{ij})  = (\gt^{ij})^{-1}$ and $\Db_k$ is the $\hb_{ij}$ covariant derivative.
Using this formula, the Einstein tensor $\Gb^{ij}$ of the metric $\gb_{ij}$
is given by
\lalign{Gb1Z}{
|\gb|\Gb^{ij} = \frac{\ep^2}{2}|\hb| \bigl[
\gt^{kl}& \Db_k\Db_l \ub^{ij} + \ep^2\bigl(a_1^{ij}+ a_2^{ij}+ a_3^{ij}\bigr) + b^{ij} + \ep c_1^{ij} + \ep^2
c_2^{ij} +
4\ep^2 \Tcb^{ij}
\bigr], \label{Gb1Z.1}
}
where
\lalign{Gb2}{
a_1^{ij} & = \Half \bigl(\Half \gt_{k l} \gt_{mn} - \gt_{km}
\gt_{l n} \bigr) \bigl(\gt^{ip} \gt^{jq} - \Half \gt^{ij}
\gt^{pq} \bigr)\Db_{p} \ub^{k l}\Db_{q} \ub^{mn}
\label{Gb2.1} ,\\
a_2^{ij} & = 2\gt_{k l}\Bigl( \gt^{n(i }
\Db_{m} \ub^{j) l}\Db_{n} \ub^{k m} - \Half
\gt^{ij}\Db_{m} \ub^{k n} \Db_{n} \ub^{m l}
- \gt^{mn}\Db_{m}  \ub^{ik}\Db_{n} \ub^{j l}\Bigr)\label{Gb2.2}, \\
a_3^{ij} & = \Db_k\ub^{ij}\Db_{l} \ub^{k l}-
\Db_k\ub^{i l}\Db_l \ub^{jk}\label{Gb2.3}, \\
b^{ij} & =  \gt^{ij}\Db_k\Db_l \ub^{k l} -
2\Db_{l}\Db_k \ub^{k(i} \gt^{j)l} \label{Gb2.4}, \\
c_1^{ij} & = -\bigl(\hb^{ij}\ep\ub^{kl} + \ep\ub^{ij}\hb^{kl}\bigr)\frac{1}{\ep^2}\Rcb_{kl}+\frac{2}{\ep^2}\Rcb_{lkm}{}^{(i}\ep\ub^{j)k}\hb^{lm}+ \frac{2}{\ep^2}\Rcb_{lkm}{}^{(i}\hb^{j)k}\ep\ub^{lm}, \label{Gb2.5} \\
c_2^{ij} & = -\ep\ub^{ij}\ep\ub^{kl}\frac{1}{\ep^2}\Rcb_{kl}+\frac{2}{\ep^2}\Rcb_{lkm}{}^{(i}\ep\ub^{j)k}\ep \ub^{lm},\label{Gb2.6}
\intertext{and}
\Tcb^{ij} & = \mu \xib^{i}\xib^{j}, \qquad \xib^{i} = \frac{1}{\ep} \delta^{i}_0. \label{Gb2.7}
}
For later use, we define
\lalign{cdefA}{
c^{ij}_{kl} = -\frac{1}{\ep^2}\Rcb_{kl}\hb^{ij} - &  \frac{1}{\ep^2}\Rcb\delta^i_k\delta^j_l + \frac{2}{\ep^2}\Rcb_k{}^{(i}\delta^{j)}_l + \frac{2}{\ep^2}\Rcb_{kml}{}^{(i}\hb^{j)m}
}
so that
\eqn{cdefB}{
c_1^{ij} = c^{ij}_{kl}\ep\ub^{kl}.
}

To fix the gauge, we set
\leqn{harm1}{
\Db_i\ub^{ij} = 0.
}
For $\ep>0$, it is clear from \eqref{ubdef} that this is equivalent to
\eqn{harm2}{
\Db_i\bigl(\sqrt{|\gb|}\gb^{ij}\bigr)=0,
}
and this is easily seen to be equivalent to the harmonic coordinate condition
\eqn{harm3}{
\gb^{ij}\bigl(\Gammab^k_{ij}-\gammab^k_{ij}\bigr) = 0.
}
Defining the reduced Einstein tensor $\Gb_R^{ij}$ by
\lalign{redGbZ}{
\Gb_R^{ij} =& \frac{1}{\ep^2}\frac{|\gb|}{|\hb|}\Gb^{ij} - b^{ij} = \Half\Bigl(\gt^{kl}\Db_k\Db_l \ub^{ij}
 + \ep^2\bigl(a_1^{ij} + a_2^{ij}+  a_3^{ij}\bigr)+ \ep c_1^{ij} + \ep^2 c_2^{ij} +
4\ep^2 \Tcb^{ij} \Bigr) \label{redGb} ,
}
Einstein's equations $\Gb^{ij}=2\ep^4 \Tb^{ij}$ in the gauge \eqref{harm1} become
\leqn{redeqns}{
\Gb_R^{ij} = 2\ep^2 \frac{|\gb|}{|\hb|}\Tb^{ij},
}
where
\leqn{stress}{
\Tb^{ij} = (\rho+\ep^2 p)\vb^i\vb^j + p \gb^{ij} \AND \vb^i\vb_j = -\frac{1}{\ep^2}.
}

To write the reduced Einstein equations \eqref{redeqns} in first order form, we introduce the variables
\leqn{udef}{
u^{ij} = \ep \ub^{ij} \AND u^{ij}_k = \Db_k \ub^{ij}.
}
With these variables, we have that
\eqn{comm1}{
\Db_k u_l^{ij} = \Db_l u_k^{ij} -\frac{2}{\ep}\Rcb_{klm}{}^{(i}u^{j)m},
}
or equivalently
\leqn{comm2}{
\delb_k u_l^{ij} = \delb_l u_k^{ij} + 2\gammab_{lm}^{(i}u_k^{j)m} -  2\gammab_{km}^{(i}u_l^{j)m}
- \frac{2}{\ep}\Rcb_{klm}{}^{(i}u^{j)m}.
}
In particular, this implies that
\leqn{comm3}{
\delb_0 u_I^{ij} = \delb_I u_0^{ij} + 2\gammab_{Im}^{(i}u_0^{j)m} -  2\gammab_{0m}^{(i}u_I^{j)m}
- \frac{2}{\ep}\Rcb_{0Im}{}^{(i}u^{j)m},
}
and hence
\lalign{comm4}{
\gt^{kl}\Db_k\Db_l \ub^{ij} &= \gt^{00}\delb_0 u_0^{ij} + \gt^{0I}\delb_0 u_I^{ij} + \gt^{0I} \delb_I u_0^{ij}
 + \gt^{IJ}\delb_I u_J^{ij} + \gt^{kl}(-\gammab_{kl}^m u_m^{ij} + 2\gammab_{km}^{(i}u_l^{j)m})\notag \\
&= \gt^{00}\delb_0 u_0^{ij} + 2\ep u^{0I}\delb_I u_0^{ij}  + \gt^{IJ} \delb_I u_J^{ij} + \ep d_1^{ij} + \ep^2 d_2^{ij},
\label{comm4.1}
}
where
\lalign{ddef}{
d_1^{ij} &= \hb^{kl}\left(-\frac{1}{\ep}\gammab_{kl}^m u_m^{ij} + \frac{2}{\ep}\gammab_{km}^{(i}u_l^{j)m}\right) \label{ddef.1}
\intertext{and}
d_2^{ij} &= u^{kl}\left(-\frac{1}{\ep}\gammab_{kl}^m u_m^{ij} + \frac{2}{\ep}\gammab_{km}^{(i}u_l^{j)m}\right)+ u^{0I}\left(\frac{2}{\ep}\gammab_{Im}^{(i}u_0^{j)m} -  \frac{2}{\ep}\gammab_{0m}^{(i}u_I^{j)m}
- \frac{2}{\ep^2}\Rcb_{0Im}{}^{(i}u^{j)m}\right).
}
Using \eqref{hcurv.1}-\eqref{hcurv.2}, we find that
\leqn{dexp}{
d_1^{ij} = d^{ijm}_{kl} u_m^{kl}
}
where
\lalign{ddefb}{
d^{ijm}_{kl} =& \frac{a'}{a}\Bigl( -\frac{3}{2} \delta^{i}_k\delta^j_l \delta^m_0 +
\delta^{(i}_0 \delta^{j)}_k \delta^I_l\delta^m_I  + \frac{1}{a} \delta^{I(i} \delta^{j)}_k \delta^m_I\delta^0_l
-\delta^{(i}_I \delta^{j)}_k \delta^I_l \delta^m_0 \Bigr). \label{ddefb.1}
}

Setting
\lalign{efdef}{
e^{ijm}_{Jkl} &=   \frac{2}{\ep} \gammab_{Jk}^{(i}\delta^{j)}_l \delta_0^m  - \frac{2}{\ep} \gammab_{0k}^{(i}\delta^{j)}_l \delta_J^m,
  \label{efdef.1} \intertext{and}
f^{ij}_{J} &= - \frac{2}{\ep^2}\Rcb_{0Jl}{}^{(i}\delta^{j)}_k, \label{efdef.2}
}
equations \eqref{comm3} and \eqref{comm4.1} can be used to write the reduced Einstein equations \eqref{redeqns} in the
following first order form
\lalign{fo1}{
-a\gt^{00}\del_t u_0^{ij} &= 2a u^{0I}\del_I u_0^{ij} + \frac{1}{\ep}a \gt^{IJ}\del_I u_J^{ij}
 + a \bigl( c^{ij}_{kl}u^{kl}  +  d^{ijm}_{kl} u_m^{kl} \bigr)  \notag \\
& \qquad + \ep a \bigl(a_1^{ij}+ a_2^{ij}+ a_3^{ij}+c_2^{ij}+d_2^{ij}\bigr) +  4\ep a \left(\Tcb^{ij}-\frac{|\gb|}{|\hb|}\Tb^{ij}\right), \label{fo1.1} \\
a \gt^{IJ}\del_t u_J^{ij} &= \frac{1}{\ep}a \gt^{IJ}\del_J u_0^{ij} + a \gt^{IJ}\bigl(e^{ijm}_{Jkl}u_m^{kl} + f^{ij}_J u^{kl} \bigr), \label{fo1.2} \\
\intertext{and}
\del_t u^{ij} &= u_0^{ij} - \frac{2}{\ep}\gammab^{(i}_{0k}u^{j)k}. \label{fo1.3}
}

To be correctly defined, the reduced Einstein equations \eqref{fo1.1}-\eqref{fo1.3} require that the matrix $\gt^{ij}$ is invertible. By assumption, $\sigma(\ep)$ is analytic in a neighborhood of
$\ep=0$ and $\sigma(0) = \sigma_0 > 0$ (see \eqref{mufix1}). Thus there exists an $\ep_0 >0$ such that\footnote{The bound  $0< \frac{\sigma_0}{2}\leq \sigma(\ep) \leq 2\sigma_0 \quad \forall \ep \in (-\ep_0,\ep_0)$ is somewhat arbitrary and could be replaced by any bound of the form $0< \frac{\sigma_0}{C}\leq \sigma(\ep) \leq C\sigma_0$. However, as we
are interested in the limit $\ep \searrow 0$, nothing is missed by assuming that $C=2$.}
\eqn{sigmaint}{
0< \frac{\sigma_0}{2}\leq \sigma(\ep) \leq 2\sigma_0 \quad \forall\, \ep  \in (-\ep_0,\ep_0).
}
For fixed
 $-\sqrt{4/(3\sigma_0)} < \tau_0 < 0$ and $\tau_1>0$, it
is clear from \eqref{flrw3.1} that
\eqn{abounds}{
\left(\sqrt{\frac{3\mu_0}{4}}\tau_0+1\right)^{\frac{4}{3}} \leq a_\ep(t) \leq \bigl(\sqrt{3\sigma_0}\tau_1+1\bigr)^{\frac{4}{3}}
}
for all $(t,\ep)\in [\tau_0,\tau_1]\times(-\ep_0,\ep_0)$.
This implies that the set
\eqn{Vcaldef}{
\Vc = \{(r^{ij})| \det(\hb^{ij}+r^{ij})>0 \, \forall\, (t,\ep)\in [\tau_0,\tau_1]\times(-\ep_0,\ep_0) \}
}
is open and  contains the origin $(r^{ij})=0$, and moreover that
the reduced Einstein equations \eqref{fo1.1}-\eqref{fo1.3} are well defined
for all $t\in (\tau_0,\tau_1)$, $\ep \in (0,\ep_0)$, and $(\ep u^{ij})\in \Vc$.
We also note that
\lalign{sdiff1Z}{
4\ep^2\left(\frac{|\gb|}{|\hb|}\Tb^{ij}-\Tcb^{ij}\right) = 4&(\rho-\mu)\delta^{i}_0\delta^{j}_0
 \notag \\ &+ \ep \bigl[ 8\delta^{(i}_0 \rho w^{j)} + 4\delta^i_0\delta^j_0 \rho \hb_{kl}u^{kl} \bigr] + \ep^2 S_0^{ij}(\ep,\sigma,u^{kl},\rho,w^k),
\label{sdiff1Z.1}}
where the map $S_0^{ij}$ is
analytic in all variables provided that
$(\ep,\sigma)\in (-\ep_0,\ep_0)\times (\sigma_0/2,2\sigma_0)$, and
$(\ep u^{ij}) \in \Vc$.

\subsect{eul}{Regularized Euler equations}

In the coordinates $(\xb^i)$, the Euler equations are given by
\leqn{eul1}{ \nablab_i \Tb^{ij} =0} where $\Tb^{ij} = (\rho + \ep^2
p) \vb^i\vb^j + p \gb^{ij}$ and the fluid velocity $\vb^i$ is
normalized according to
\leqn{eul2}{\vb_i\vb^i =
-\frac{1}{\ep^2}\, .}
To derive a symmetric hyperbolic system for the Euler system, we follow the method of
\cite{BK07} and differentiate \eqref{eul2} to get
\leqn{eul3}{\vb_i \nablab_j \vb^i = 0}
which in turn implies
\leqn{eul4}{\vb^{j}\vb_i \nablab_j \vb^i = 0.}
Writing out
\eqref{eul1} explicitly, we have
\leqn{eul5}{(\delb_i\rho
+\ep^2\delb_i p)\vb^i\vb^j + (\rho+\ep^2 p)(\vb^j\nablab_i\vb^i
+\vb^i\nablab_i\vb^j) + \gb^{ij}\delb_i p = 0\, . }
Next, we observe that the operator
\eqn{proj}{L^j_i = \delta^j_i + \ep^2 \vb^j\vb_i}
projects into
subspace orthogonal to the fluid velocity $\vb^i$, i.e.
$L^j_{i}L^i_k = L^j_k $ and $L^j_i\vb^i=0$.
Applying this operator to project \eqref{eul5} into components parallel
and orthogonal to $\vb^i$ yields, after using the relations
\eqref{eul2}-\eqref{eul4}, the following system
\lgath{eul6}{
\vb^i\delb_i
\rho +(\rho+\ep^2 p)L^i_j\nablab_i\vb^j = 0 ,\label{eul6.1} \\
M_{ij}\vb^k\nablab_k \vb^j + \frac{1}{\rho+\ep^2 p} L^i_j \delb_i
p = 0,\label{eul6.2}
}
where
\eqn{Mdef}{ M_{ij} = \gb_{ij} +
2\ep^2 \vb_i\vb_j .
}
Changing to the Makino density variable $\alpha$ (see \eqref{dendef}) and multiplying \eqref{eul6.1} by the square of the function
\eqn{fdef}{
f = \left( 1 + \frac{1}{4n(n+1)}(\ep\alpha)^2\right)\,
}
transforms the Euler equations \eqref{eul6.1}-\eqref{eul6.2} into
\lgath{eul7}{
f^2 \vb^i\delb_i
\alpha + \frac{\alpha}{2nf} L^i_j\nablab_i\vb^j = 0 ,\label{eul7.1} \\
M_{ij}\vb^k\nablab_k \vb^j + \frac{\alpha}{2nf} L^i_j \delb_i
\alpha = 0 .\label{eul7.2}
}

As discussed in the introduction, we need to introduce a new fluid four-vector
by
\leqn{wdef}{
w^i = \vb^i-\xib^i = \vb^i-\frac{\delta^i_0}{\ep}.
}
So, letting
\leqn{wvdef}{
W_M = (\alpha,w^i)^T
}
allows us to write the system \eqref{eul7.1}-\eqref{eul7.2} as
\leqn{eul7}{
A_M^0\del_t W_M = A_M^I\del_I W_M + F_M,
}
where
\lalign{eul8}{
A_M^0 & = \begin{pmatrix} f^2(1+\ep w^0) &  \frac{\ep \alpha}{2nf}  L^0_j \\
\frac{\ep \alpha}{2nf} L^0_i &  M_{ij}(1+\ep w^0) \end{pmatrix}, \label{eul8.1} \\
A_M^I & = \begin{pmatrix} - f^2 w^I &  -\frac{\alpha}{2nf} L^I_j \\
-\frac{\alpha}{2nf} L^I_i &  -M_{ij}w^I \end{pmatrix}, \label{eul8.2}
\intertext{and}
F_M & = \begin{pmatrix} \frac{\alpha}{2nf} L^i_j\bigl(\gammab^{j}_{il}-\Gammab^{j}_{il}\bigr)\vb^l
-\frac{\alpha}{2nf} L^i_j\gammab^{j}_{il}\vb^l \\
M_{ij}\bigl(\gammab^{j}_{kl}-\Gammab^{j}_{kl}\bigr)\vb^k\vb^l - M_{ij} \gammab^{j}_{kl} \vb^k\vb^l
\end{pmatrix}. \label{eul8.3}
}
Next, a straightforward calculation using \eqref{utog} and \eqref{christ} shows that
\alin{eul9}{
& M_{ij}  = \hb_{ij} + 2\delta_i^0\delta_j^0 + \ep m_{ij}(\ep,\sigma,u^{kl},w^k), \\
&L^{i}_{j}  = \delta^{i}_j - \delta^i_0\delta^0_j + \ep\ell_i^j(\ep,\sigma,u^{kl},w^k), \\
& L^i_j\bigl(\gammab^{j}_{il}-\Gammab^{j}_{il}\bigr)\vb^l = \ep q_0(\ep,\sigma,u^{ij},u^{ij}_k,w^i), \\
& L^i_j\gammab^{j}_{il}\vb^l = \frac{3}{2}\frac{a'}{a}
+ \ep q_1(\ep,\sigma,u^{ij},w^i),\\
& \bigl(\gammab^{j}_{kl}-\Gammab^{j}_{kl}\bigr)\vb^k\vb^l = -u^{j0}_0 + \frac{1}{4}\delta^{j0}(3 u^{00}_0 - a\delta_{KL}u^{KL}_0) \\
&\text{\hspace{3.0cm}}
-\frac{1}{4}\delta^{jI}\left(\frac{1}{a} u_I^{00} + \delta_{KL} u^{KL}_I\right) + \ep q^j_0(\ep,\sigma,u^{ij},u^{ij}_k,w^i),\\
& \gammab^{j}_{kl} \vb^k\vb^l = \frac{a'}{a}\delta^j_I w^I + \ep \left[ \frac{1}{\ep} \gammab^j_{kl}w^k w^l\right],
}
where the maps $m_{ij}$, $\ell^j_i$, $q_0$, $q_1$, and $q_0^j$ are analytic in all their variables
provided that $(\ep,\sigma) \in (-\ep_0,\ep_0)\times (\sigma_0/2,2\sigma)$ and $(\ep u^{ij})\in \Vc$.
Using these expressions, we can decompose $A^0_M$, $A^I_M$, and $F_M$ as
\lalign{AFdec1}{
A^0_M &= A^0_{M,0} + \ep A^0_{M,1}(\ep,\sigma,u^{ij},\alpha,w^i), \label{AFdec1.1} \\
A^I_M &= A^I_{M,0}+\ep A^I_{M,1}(\ep,\sigma,u^{ij},\alpha,w^i), \label{AFdec1.2} \\
F_M & = F_{M,0} + \ep F_{M,1}(\ep,\sigma,u^{ij},u^{ij}_k,\alpha,w^i), \label{AFdec1.3}
}
where
\lalign{AFdec2}{
A^0_{M,0} & = \begin{pmatrix} 1 & 0 \\ 0 & \hb_{ij}+ 2\delta_i^0\delta_j^0\end{pmatrix}, \label{AFdec2.1} \\
A^I_{M,0} & = \begin{pmatrix} - w^I & -\frac{\alpha}{2n} \delta^I_j \\
                              -\frac{\alpha}{2n} \delta^I_i & -( \hb_{ij}+ 2 \delta^0_i\delta^0_j)w^I \end{pmatrix},
                              \label{AFdec2.2} \\
F_{M,0} & = \begin{pmatrix} -\frac{3a'}{4na}\alpha  \\
( \hb_{ij}+ 2\delta^0_i\delta^0_j)\left[-u^{j0}_0 + \frac{1}{4}\delta^{j0}(3 u^{00}_0 -
a\delta_{KL}u^{KL}_0)
-\frac{1}{4}\delta^{jI}\left(\frac{1}{a} u_I^{00} + \delta_{KL} u^{KL}_I\right) - \frac{a'}{a}\delta^j_I w^I \right]\end{pmatrix}, \label{AFdec2.3}
}
and the maps $A^0_{M,1}$, $A^I_{M,0}$, $F_{M,1}$ are analytic in all their variables
provided that  $(\ep,\sigma)\in (-\ep_0,\ep_0)\times (\sigma_0/2,2\sigma_0)$ and $(\ep u^{ij})\in \Vc$.

\subsect{nloc}{A nonlocal symmetric hyperbolic formulation}

To bring the reduced Einstein equations \eqref{fo1.1}-\eqref{fo1.3} into a form that is suitable to analyze the
limit $\ep\searrow 0$, we replace the $u^{ij}_J$ with the variables
\leqn{Wdef1}{
W^{ij}_I = u^{ij}_I - \delta^i_0\delta^j_0\del_I\Phi
}
where $\Phi$ satisfies the Poisson equation
\leqn{Phidef}{
\Delta \Phi =  4 a \Pi(\rho-\mu).
}
where $\Delta = \delta^{IJ}\del_I\del_J$ is the flat Laplacian.
In addition to $\Phi$, we also need the time derivative
\leqn{Phitdef}{
\Phid = \del_t\Phi
}
which satisfies
\leqn{Phidef1Z}{
\Delta \Phid = 4a \Pi\bigl(\del_t(\rho-\mu)\bigr) + 4a'\Pi\bigl(\rho-\mu\bigr).
}
Using \eqref{eul6.1}-\eqref{eul6.2} to replace the time derivatives of $\rho$ and $\del_t w^i$ in favor of
spatial derivatives, we find that
\lalign{rhodot}{
\del_t(\rho-\mu) = -\frac{3}{2}\frac{a'}{a}(\rho-\mu)& - \del_I(\rho w^I) + \ep \left[
\del_I(\rho w^I w^0) + \rho\left(\frac{3}{4} u^{00}_0
-\frac{a}{4} \delta_{IJ} u^{IJ}_0\right)  \right] \notag \\
& -\ep^2 \alpha^2\Bigl[Q_0(\ep,\sigma,u^{ij},\alpha,w^i)+ Q_1(\ep,\sigma,u^{ij},\alpha,w^i,\del_I \alpha,\del_I w^i) \Bigr],
\label{rhodot.1}}
where $Q_\nu$ $(\nu=0,1)$ are analytic in all variables for $(\ep,\sigma)\in (-\ep_0,\ep_0)\times (\sigma_0/2,2\sigma_0)$, and
$(\ep u^{ij}) \in \Vc$, and $Q_1$ is linear in $(\del_I \alpha,\del_I w^i)$.

Letting,
\leqn{phidef}{
\phi = -\frac{4}{\ep}\ip{\oh}{\rho-\mu}_{L^2},
}
we see from \eqref{rhodot.1} that $\phi$ satisfies
\lalign{phidot}{
\phi' = -&\frac{3}{2}\frac{a'}{a}\phi - 4 \Bigl[
\ipB{\rho}{\frac{3}{4} u^{00}_0
-\frac{a}{4} \delta_{IJ} u^{IJ}_0}_{L^2} \Bigr] +  4 \ep \ip{\alpha^2 \oh}{Q_0+ Q_1}_{L^2}.
\label{phidot.1}
}
Define
\leqn{Wdef}{
W = (u_0^{ij},W_I^{ij},u^{ij},\phi^{ij},\alpha,w^j)^T,
}
\eqref{fo1.1}-\eqref{fo1.3}, \eqref{sdiff1Z.1},
\eqref{eul7}, \eqref{eul8.1}-\eqref{eul8.3}, \eqref{AFdec1.1}-\eqref{AFdec1.3}, \eqref{AFdec2.2}-\eqref{AFdec2.3},
 \eqref{Wdef1}, \eqref{phidef} and \eqref{phidot.1} that $W$ satisfies
\leqn{nonloc1}{
A^0 \del_t W = \frac{1}{\ep}C^{I}\del_I W + (A_0^I + \ep A_1^I)\del_I W + F_0 + \ep F_1,
}
where
\lalign{Adef}{
A^0 & = \begin{pmatrix} A^0_G & 0 \\ 0 & A^0_{M,0}+\ep A^0_{M,1} \end{pmatrix} \label{Adef.1} , \\
A^0_G & = \begin{pmatrix} a(1-\ep u^{00}) & 0 & 0 & 0\\
0  & (\delta^{IJ}+\ep a u^{IJ})  & 0  & 0\\
0 & 0 & 1  & 0 \\
0 & 0 & 0 & 1 \end{pmatrix} \label{Adef.2} ,
}
\lalign{Adef1}{
A^I_0 & = \begin{pmatrix} A^I_G & 0 \\ 0 & A^I_{M,0} \end{pmatrix} \label{Adef.3} , \\
A^I_G & = \begin{pmatrix} 2a u^{0I} & a u^{IJ} & 0 & 0 \\
a u^{IJ} & 0 & 0 & 0\\
0 & 0 & 0 & 0 \\
0 & 0 & 0 & 0 \end{pmatrix} \label{Adef.4},  \\
A^I_1 & = \begin{pmatrix} 0 & 0 \\ 0 & A^I_{M,1} \end{pmatrix} \label{Adef.5} ,
}
\lalign{Adef2}{
C^I & = \begin{pmatrix} C^I_G & 0 \\ 0 & 0 \end{pmatrix} \label{Adef.6},\\
C^I_G & = \begin{pmatrix} 0 & \delta^{IJ} & 0 & 0 \\ \delta^{IJ} & 0 & 0 & 0 \\
0 & 0 & 0 & 0 \\
0 & 0 & 0 & 0 \end{pmatrix}, \label{Adef.7}
}
and
\lalign{Fdef}{
F_0 & = \begin{pmatrix} a\Bigl( \delta^i_0\delta^j_0\bigl(u^{IJ}\del_I\del_J\Phi + \phi - 4\rho \hb_{kl}u^{kl}\bigr)
                                 -8 \delta_0^{(i}\rho w^{j)} + c^{ij}_{kl}u^{kl} + d^{ijm}_{kl} u_m^{kl}\Bigr) \\
                        -\delta^{IJ}\del_J \Phid_0 \delta^i_0\delta^j_0 + \delta^{IJ}\bigl(e^{ijm}_{Jkl}u_m^{kl} + f^{ij}_J u^{kl} \bigr) \\
                        u_0^{ij} - \frac{2}{\ep}\gammab^{(i}_{0k}u^{j)k}  \\
                              -\frac{3}{2}\frac{a'}{a}\phi - \ipb{\rho}{3u^{00}_0
-a\delta_{IJ} u^{IJ}_0}_{L^2} \\
                           -\frac{3a'}{4na}\alpha \\
( \hb_{ij}+ 2\delta^0_i\delta^0_j) \Bigl(-u^{j0}_0 + \frac{1}{4}\delta^{j0}(3 u^{00}_0 -
a\delta_{KL}u^{KL}_0) - \frac{1}{4}\delta^{jI}\Bigl(\frac{1}{a}u_I^{00} +
\delta_{KL} u^{KL}_I\Bigr) - \frac{a'(t)}{a}\delta^j_I w^I\Bigr)
                             \end{pmatrix}, \label{Fdef.1} \\
\Phid_0  & = -4 a\Pi\left(\del_I(\rho w^I) + \frac{1}{2}\frac{a'}{a} (\rho-\mu)\right),
\label{Fdef.2} \\
F_1 & = \begin{pmatrix}  a\bigl(a_1^{ij}+ a_2^{ij}+ a_3^{ij}+c_2^{ij}+d_2^{ij} - S_0^{ij} \bigr)  \\
-\frac{1}{\ep}\delta^{IJ}\del_J(\Phid-\Phid_0)\delta^i_0\delta^j_0  + a u^{IJ}\bigl[\del_J\Phid \delta^i_0\delta^j_0+ e^{ijm}_{Jkl}u_m^{kl} + f^{ij}_J u^{kl}\bigr] \\
          4\ip{\alpha^2 \oh}{Q_0+Q_1}   \\
          \frac{1}{\ep}\bigl(F_{M,0}-\tilde{F}_{M,0}\bigr)+ F_{M,1} \end{pmatrix}. \label{Fdef.3}
}
For $\ep>0$, equation \eqref{nonloc1} is completely equivalent to the Einstein-Euler equations in the harmonic gauge. It is this form
of the Einstein-Euler equations that will be useful for analyzing the limit $\ep \searrow 0$.

\subsect{welldef}{Well-posedness of the nonlocal system}

The well-posedness of the non-local symmetric hyperbolic system \eqref{Wdef} follows from the same arguments used in Section 2.4 of \cite{Oli09}, and will
not be repeated here. The well-posedness of the system \eqref{Wdef} combined with its particular structure allows us to apply the local existence results of Schochet \cite{Scho86,Scho88} (see also \cite{KM82,Kreiss80}) to
obtain the existence of one-parameter families  of solutions to \eqref{Wdef}
on spacetime regions of the form $M=[0,T)\times \Tbb^3$ where $T$ is independent of $\ep$, and also to identify the cosmological Poisson-Euler equations
\eqref{limB.1}-\eqref{limB.3} as the correct limit equations satisfied by the $\ep \searrow 0$ limit of the solutions of \eqref{Wdef}. The details
of this are presented in Sections \ref{limit} and \ref{nlim}. 

%% file: idata.tex
\sect{idata}{Newtonian initial data}

In order to solve the initial value problem for the Einstein equations, we must first construct initial data that satisfies the
following constraint equations on the initial hypersurface defined by $t=0$:
\lalign{ceqns}{
\bigl(\Gb^{0j}-2\ep^4 \Tb^{0j}\bigr)\bigl|_{t=0} &= 0, \label{ceqns.1} \\
 \Db_i \ub^{ij}\bigl|_{t=0} &= 0,\label{ceqns.2}
\intertext{and}
\left(\gb_{ij}\vb^{i} \vb^{j} + \frac{1}{\ep^2}\right)\Bigl|_{t=0} &= 0,
\label{ceqns.3}
}
which are the gravitational constraints, the harmonic gauge condition, and the fluid 4-velocity normalization, respectively.

To find 1-parameter families of solutions to these equations, we adapt the method developed by Lottermoser in \cite{Lott}.
We begin by writing \eqref{ceqns.2} as (see \eqref{udef})
\leqn{harmZ}{
u^{kj}_j\bigl|_{t=0} = 0,
}
which shows that
\leqn{dharmZ}{
\del_J u^{kj}_k \bigl|_{t=0} = 0,
}
and hence, that
\leqn{DharmZ}{
\Db_i u^{kj}_k \bigl|_{t=0} = \delta_i^0 \ep\del_t  u^{kj}_k .
}
Substituting this into \eqref{Gb2.4} yields
\leqn{btez}{
b^{0j}\bigl|_{t=0} = - \gh^{00}\ep\del_t u^{jk}_k.
}
Setting
\eqn{Hdef}{
H^{ij} = \bigl(\gh^{kl}\Db_k\Db_l \ub^{ij} + b^{ij}\bigr)|_{t=0},
}
it follows directly from \eqref{comm4.1} and \eqref{btez} that
\leqn{Hdefj}{
H^{0j} = \gh^{IJ} \del_I u_J^{0j} - \gh^{00}\ep\del_t u^{jI}_I + 2\ep u^{0I}\del_I u^{0j}_0
+\ep d_1^{0j} + \ep^2 d^{0j}_2.
}
Setting $j=J$ in the above expression while using \eqref{harmZ} gives
\leqn{HJ}{
H^{0J} = \gh^{KL} \del_K u_L^{0J} - \gh^{00}\ep\del_t u^{JI}_I - 2\ep u^{0I}\del_I u^{KJ}_K
+\ep d_1^{0J} + \ep^2 d^{0J}_2.
}
From the non-vanishing Christoffel symbols \eqref{hcurv.1}-\eqref{hcurv.2}, we get that
\lalign{DuexpA}{
u^{00}_0 &= \ep \del_t \ub^{00}, \label{DuexpA.1} \\
u^{0J}_0  & = \ep \del_t \ub^{0J} + \frac{a'}{2a} u^{0J}, \label{DuexpA.2} \\
u^{IJ}_0 & = \ep \del_t \ub^{IJ} + \frac{a'}{a} u^{IJ}, \label{DuexpA.3}
}
and
\lalign{DuexpB}{
u^{00}_K &= \del_K \ub^{00} + a' u^{0K}, \label{DuexpB.1} \\
u^{0J}_K & = \del_K \ub^{0J} + \frac{a'}{2}u^{KJ}+ \frac{a'}{2a} \delta^J_K u^{00}, \label{DuexpB.2}\\
u^{IJ}_K & = \del_K \ub^{IJ} + \frac{a'}{a} \delta^{(I}_K u^{J)0}. \label{DuexpB.3}
}
Using \eqref{hcurv.3}-\eqref{hcurv.4}, \eqref{harmZ}, and \eqref{DuexpA.1}-\eqref{DuexpB.3}, a straightforward calculation (see \eqref{Gb2.5} and \eqref{ddef.1}) shows
that
\lalign{dpc}{
\ep \del_t u^{IJ}_I & + \ep(d_1^{0J}+c^{0J}) = \del_I(\ep\del_t \ub^{IJ})+ \ep \left[ -\frac{3}{2}\left(\frac{a'}{a}\right)^2 u^{J0} + \frac{a'}{2a} \del_I\ub^{JI}+ \frac{a'}{2a^2}\del_J \ub^{00}\right]. \label{dpc.1}
}
Setting $j=0$ in \eqref{Hdefj} while using \eqref{comm3} and \eqref{harmZ} gives
\lalign{HO}{
H^{00} = \gh^{KL}& \del_K u_L^{00} -\del_I u_J^{IJ}  + \ep\Bigl[
u^{00}\del_I u_J^{IJ} - \gh^{00}\Bigl( \frac{2}{\ep} \gammab^{(0}_{0m} u^{I)m}_I \notag\\
&- \frac{2}{\ep} \gammab^{(0}_{Im} u^{I)m}_0
- \frac{2}{\ep^2} \Rcb_{0Im}{}^{(0}u^{I)m}\Bigr) - 2 u^{0I}\del_I u_J^{J0} + d_1^{00} + \ep d_2^{00} \Bigr] \label{HO.1}
}

Next, we decompose the gravitational variables as follows
\lalign{gvd1}{
\ub^{0j} &= \frac{1}{\ep}y^{I}\delta_I^j + \uc^{0j},\label{gvd1.1}\\
\ub^{IJ} &= \ep \uc^{IJ}, \label{gvd1.2}\\
\del_t\ub^{IJ} & = \uc^{IJ}_0, \label{gvd1.3}
}

where we assume that the $y^J$ are constants and
\leqn{gvd2}{
\int_{\Tbb^3} \uc^{0j}\, d^3 x = 0.
}
In terms of these variables, the harmonic conditions become
\lalign{harmuc}{
\del_t \ub^{00} &= -\frac{1}{\ep}\del_I \uc^{I0} - \frac{3}{2}\ln(a)' \uc^{00} -\frac{1}{2} a' \delta_{IJ} \ep \uc^{IJ}, \label{harmuc.1}\\
\del_t \ub^{0J} & = -\del_I\uc^{IJ} - \frac{5}{2}\ln(a)'
\left(\frac{1}{\ep}y^{J} + \uc^{0J} \right). \label{harmuc.2}
}
Using \eqref{Gb1Z.1}, \eqref{HJ}, \eqref{dpc.1}, \eqref{HO.1}, \eqref{harmuc.1}, and \eqref{harmuc.2}, the gravitational constraints
\eqref{ceqns.1} take the form
\lalign{gravc}{
&\Delta \uc^{00} - 4a(\rho-\mu) + \ep \bigl[f^{0}_1(\ep,\sigma,y,\uc,\rho,w)
+f^0_1(\ep,\sigma,y,\uc,\del^2_x\uc,\del_x \uc,\uc_0)
+ f^0_2(\ep,\sigma,y,\uc,\del_x \uc,\uc_0)\bigr] = 0,\label{gravc.1} \\
&\Delta\uc^{0J} + \ep a \Bigl[\frac{a'}{a}\del^J\uc^{00} + \del_I\uc^{IJ}_0 +  y^{KL}\del_K\del_L\uc^{0J}\Bigr]
+ \ep a \Bigl[-\frac{3}{2}\Bigl(\frac{a'}{a}\Bigr)^2 y^{J} - 4\rho w^J \Bigr]
\notag \\
& \text{\hspace{2.0cm}} + \ep^2 \Bigl[f^{J}_1(\ep,\sigma,y,\uc,\rho,w)
+f^J_1(\ep,\sigma,y,\uc,\del^2_x\uc,\del_x \uc,\uc_0)
+ f^J_2(\ep,\sigma,y,\uc,\del_x \uc,\uc_0) \Bigr] = 0, \label{gravc.2}
}
where \textbf{(i)}
for any $R>0$ there exists an $\ep_0 > 0$ such that the maps $f_\alpha$ $(\alpha=2,3,4)$
are analytic in all their variables provided $|\ep|<\ep_0$, $\sigma\in (\sigma_0/2,2\sigma_0)$ and $|\ub^{ij}|<R$, \textbf{(ii)}
$ f^j_{2}$ is linear in $(\del_K\del_L\ub^{ij},\del_K\ub^{ij},\del_t\ub^{ij})$, and \textbf{(iii)} $f^j_2$  is quadratic in
$(\del_t\ub^{ij},\del_K \ub^{ij})$. Here, $a$, $a'$, and $\mu$ are determined
by evaluating the formulas \eqref{flrw1.3}, \eqref{flrw3.1} and \eqref{flrw3.2} at $t=0$, that is
\leqn{apsigma}{
a = 1, \quad \mu = \sigma, \AND a' = \sqrt{\frac{8}{3}\sigma}.
}
Also, a calculation using \eqref{utog} shows that \eqref{ceqns.3} can be written as
\leqn{ceqn3}{
 w^0 - \ep f_0\bigl(\ep,\sigma,y,\uc^{ij},w^{I}\bigr) = 0,
}
where the map $f_0$ is
analytic provided $|\ep|<\ep_0$, $\sigma\in (\sigma_0/2,2\sigma_0)$, $|\ub^{ij}|<R$, and $|w^I|<R$.

\begin{thm} \label{idataA} \mnote{[idataA]}
Suppose $s>3/2$, $R>0$, $\alpha \in H^{s}$, $w^I\in H^{s}(\Rbb^3)$,
$\uc^{IJ}_0 \in \Hb^{s}(\Sbb{3})$ , $\uc^{IJ} \in B_R\bigl(\Hb^{s+1}(\Sbb{3})\bigr)$. Then there exists an
$\ep_0 > 0$, and
analytic maps
\alin{idataA1}{
&(-\ep_0,\ep_0) \ni \ep \longmapsto \sigma(\ep) \in \Rbb_{>0},\\
&(-\ep_0,\ep_0) \ni \ep \longmapsto w^0_\ep \in H^s, \\
&(-\ep_0,\ep_0) \ni \ep \longmapsto y^J(\ep) \in \Rbb^3,
\intertext{and}
&(-\ep_0,\ep_0) \ni \ep \longmapsto \uc^{0j}_\ep=\uc^{j0}_\ep \in \Hb^s(\Rbb^4), \\
}
such that for every $\ep \in (-\ep_0,\ep_0)$

\alin{idataA2}{
\alpha|_{t=0} & = \alpha, \\
(w^j_\ep)|_{t=0} &= (w^0_\ep,w^I), \\
(\ub^{ij}_\ep)\bigl|_{t=0} & = \begin{pmatrix} \uc^{00}_\ep & \frac{1}{\ep} y^J(\ep) + \ub^{0J}_\ep \\
\frac{1}{\ep}y^J(\ep) + \uc^{J0}_\ep & \ep \uc^{IJ} \end{pmatrix}
\intertext{and}
(\del_t\ub^{ij}_\ep)\bigl|_{t=0} &= \begin{pmatrix} \del_t\ub^{00}_\ep\bigl|_{t=0}  & \del_t\ub^{0J}_\ep|_{t=0} \\
\del_t\ub^{0J}_\ep\bigl|_{t=0} & \uc_0^{IJ} \end{pmatrix}
}
where
\eqn{idataA3a}{
\del_t\ub^{00}_\ep\bigl|_{t=0} = -\frac{1}{\ep}\del_I \uc^{I0}_\ep - \frac{3}{2}\ln(a)' \uc^{00}_\ep -\frac{1}{2} a' \delta_{IJ} \ep \uc^{IJ},
}
and
\eqn{idataA3b}{
\del_t\ub^{0J}_\ep\bigl|_{t=0} = -\del_I\uc^{IJ} - \frac{5}{2}\ln(a)'
\left(\frac{1}{\ep}y^{J}(\ep) + \uc^{0J}_\ep\right),
}
satisfy the gravitational constraint equations \eqref{ceqns.1}, the harmonic gauge condition \eqref{ceqns.2}, and the fluid velocity normalization \eqref{ceqns.3}.
Moreover, the maps
$\sigma(\ep)$, $w^0_\ep$, $y^J(\ep)$ and $\uc^{0j}_\ep$ admit the expansions
\gath{idataA4}{
\sigma(\ep)  = \sigma_0 + \text{\rm O}(\ep), \quad
w^0_\ep = \text{\rm O}(\ep), \quad
y^J(\ep) =  y^J_0 + \text{\rm O}(\ep), \\
\uc^{00}_\ep =  \phi + \text{\rm O}(\ep),
\AND
\uc^{0J}_\ep  = \text{\rm O}(\ep),
}
where
\gath{idataA5}{
\sigma_0 = \int_{\Tbb^3} \rho \, d^3 x,\quad
y^J_0  = - \frac{1}{\sigma_0}\int_{\Tbb^3}\rho w^J \, d^3 x,
\intertext{and}
\phi = \Delta^{-1}(\rho-\sigma_0).
}
\end{thm}
\begin{proof}
Applying the projection operators $\Pi$ and $\id-\Pi$ to the equations \eqref{gravc.1}-\eqref{gravc.2}, while observing  the averaging
conditions \eqref{gvd2} and the definitions \eqref{apsigma}, yields the equations
\lalign{gravc2}{
&\sigma - \ip{1}{\rho}_{L^2} + \frac{\ep}{4}\ip{1}{F^0(\ep,\sigma,y,\uc,\uc_0,\alpha,w)}_{L^2}  = 0, \label{gravc2.1} \\
&y^J + \frac{1}{\sigma}\ip{1}{\rho w^J}_{L^2} -\frac{1}{4\sigma}\ep\ip{1}{F^J(\ep,\sigma,y,\uc,\uc_0,\alpha,w )}_{L^2} =  0,\label{gravc2.2}\\
&\Delta \uc^{00} -4\Pi(\rho) + \ep \Pi\bigl(F^0(\ep,\sigma,y,\uc,\uc_0,\alpha,w )\bigr)= 0, \label{gravc2.3}
\intertext{and}
&\Delta \uc^{0J} + \ep \Pi\bigl(-4\rho w^J + \ep F^J(\ep,\sigma,y,\uc,\uc_0,\alpha,w)\bigr) = 0 \label{gravc2.4}.
}
These equations together with \eqref{ceqn3}
form the complete set of constraint equations to be solved

Using the same arguments as in Section 3 of \cite{Oli09}, it can be shown that for any $s>3/2$, $R>0$, and $\sigma_0>0$ that
there exists an $\ep>0$ such that
the maps
\alin{idataA6}{
(\ep,\sigma,y^J,\uc^{ij},\uc^{IJ}_0,\alpha,w^j)& \ni (-\ep_0,\ep_0)\times (\sigma_0/2,2\sigma_0) \times \Rbb^3 \times \\
& B_R\bigl(\Hb^{s+1}(\Sbb{4})\bigr)
\times H^{s}(\Sbb{3})\times H^s\times H^s(\Rbb^4) \mapsto F^j \in H^{s-1}(\Rbb^4)
}
and
\alin{idataA7}{
(\ep,\sigma,y,\uc^{ij},w^{I}) & \ni (-\ep_0,\ep_0)\times (\sigma_0/2,2\sigma_0) \times \Rbb^3\times B_R\bigl(\Hb^{s+1}(\Sbb{4})\bigr) \times H^s(\Rbb^3) \mapsto f_0 \in H^{s}
}
are analytic. Setting
\eqn{idataA8}{
\psi = (\uc^{IJ},\uc^{IJ}_0,\alpha,w^J)^T \AND \eta = (\sigma,y^J,\uc^{0j},w^0),
}
allows us to write the constraint equations \eqref{ceqn3}, \eqref{gravc2.1}-\eqref{gravc2.4} as
the following single equation:
\leqn{idataA9}{
L(\psi,\eta) + \ep M(\ep,\psi,\eta)  = 0
}
where
\leqn{idataA10.1}{
L(\psi,\eta) =
\begin{pmatrix}\sigma - \ip{1}{\rho}_{L^2} \\ y^J + \frac{1}{\sigma}\ip{1}{\rho w^J}_{L^2} \\
\Delta \uc^{0j}-4\Pi(\rho)\delta^{0j} \\ w^0 \end{pmatrix}}
and $M(\ep,\psi,\eta)$ are analytic maps. It is clear that
\leqn{idataA11.1}{
\eta(\psi) =
\begin{pmatrix} \ip{1}{\rho}_{L^2}\\ -\frac{1}{\ip{1}{\rho}_{L^2}}\ip{1}{\rho w^J}_{L^2} \\ 4\delta^{0j}\Delta^{-1}\Pi(\rho) \\ 0\end{pmatrix}}
solves
\leqn{idataA12}{
L(\psi,\eta(\psi)) = 0.
}
Also, a straightforward calculation shows that
\leqn{idataA13.1}{
D_{2}L(\psi,\eta(\psi))\cdot \delta\eta =
\begin{pmatrix}\delta\sigma \\ \delta y^J - \frac{\ip{1}{\rho w^J}_{L^2}}{\ip{1}{\rho}_{L^2}^2}\delta\sigma \\
\Delta \delta \uc^{0j} \\ \delta w^0 \end{pmatrix},
}
and it follows easily  from this equation and the invertibility of the Laplacian
$\Delta : \Hb^{s+1} \longrightarrow \Hb^{s-1}$ that the linear map
\lalign{idataA14}{
D_{2}L(\psi,\eta(\psi)) \: : \: &\Rbb \times \Rbb^3 \times \Hb^{s+1}(\Rbb^4)\times H^s \longrightarrow \Rbb \times \Rbb^3 \times \Hb^{s-1}(\Rbb^4)\times H^s \label{idataA14.1}
}
is an isomorphism. The proof of the Theorem now follows from \eqref{idataA9}-\eqref{idataA14.1} and the analytic version
of the Implicit Function Theorem (see \cite{Deim}, Theorem 15.3).
\end{proof}

%% file: limit.tex
\sect{limit}{Limit equations}

The evolution equations \eqref{Wdef} are now in a form to
which the theory of singular symmetric hyperbolic equations \cite{Kreiss80,Scho86,Scho88} pioneered and developed  by Kreiss, Klainerman, Majda, and Schochet applies.  We know from
the theory developed in these works that the appropriate limit equation that is satisfied
by solutions of \eqref{Wdef} in the limit $\ep \searrow 0$ is
\lalign{limeqn1}{
\At^0_0 \del_t W &= \At^I_0 \del_I W + \Ft_0 + C^I \del_I\omega, \label{limeqn1.1} \\   
C^I\del_I W &= 0, \label{limeqn1.2}
}
where
\alin{limeqn2}{
\Ft_0 &= F_0\bigl|_{\ep=0}, \\
\At^0_0 &= \begin{pmatrix} \At^0_{G,0}  & 0 \\
                        0 & A_{0,M}|_{\ep=0}
        \end{pmatrix},
\intertext{and}
\At^0_{G,0} &= \begin{pmatrix} \at & 0 & 0 & 0\\
                            0 & \delta^{IJ} & 0 & 0 \\
                            0   & 0     & 1 & 0 \\
                            0 & 0 & 0 & 1
            \end{pmatrix}.
}
As we shall shortly see, solutions to the cosmological Poisson-Euler-Makino equations \eqref{limA.1}-\eqref{limA.3} determine solutions
to the limit equations \eqref{limeqn1.1}-\eqref{limeqn1.2}. But first, we record the following local existence and uniqueness result
for the Poisson-Euler-Makino equations.
\begin{prop} \label{limA} \mnote{[limA]}
Suppose $s \geq 3 + \ell$,
$\alphat_0 \in H^k$, and  $\wt^I_0 \in H^{k}(\Rbb^3)$. Then there exists a maximal time $T_0$ and
a unique solution
\gath{limA1}{
\alphat \in X_{T_0,\ell,s}, \quad \wt^I \in X_{T_0,\ell,s}(\Rbb^3), \quad \Phit \in X_{T_0,\ell+2,s+2}
}
to the Poisson-Euler-Makino equations \eqref{limA.1}-\eqref{limA.3} with initial data $\alphat|_{t=0} = \alphat_0$ and
$\wt^I|_{t=0} = \wt^I_0$.
\end{prop}
\begin{proof}
Since the Poisson-Euler-Makino equations \eqref{limA.1}-\eqref{limA.3} form a (non-local) symmetric hyperbolic system,
the proof follows from standard theory. For example, see \cite{TayIII}, Chapter 16.
\end{proof}

\begin{prop} \label{limB} \mnote{[limB]}
Suppose $\{\alphat,\wt^I,\Phit\}$ is the solution to the Poisson-Euler-Makino equations
from Proposition \ref{limA}, and let
\alin{limB2}{
\at &= \exp\left( \int_{0}^{t}\left(\frac{8}{3} \int_{\Tbb^3} \rhot(s)  d^3 x\right)^{\frac{1}{2}} ds \right), \\
\mut & = \int_{\Tbb^3} \rhot \, d^3 x, \\
\zeta^J & = \int_{\Tbb^3} \rhot \wt^J \, d^3 x, \\
\ut^{ij}_0 & =2\delta_0^{(i}\delta_J^{j)} \left( -\left(\frac{1}{\mut}\zeta^J\right)' - \frac{\at'}{2 \at}\frac{1}{\mut}\zeta^J\right), \\
\ut^{ij} & = 2\delta_0^{(i}\delta_J^{j)}\left( -\frac{1}{\mut}\zeta^J\right),\\
\Wt_I^{00} & = \at' \ut^{0I},\\
\Wt_{I}^{J0} & = 0,\\
\Wt_{I}^{JK} & = \frac{\at'}{\at}\delta^{(J}_I \ut^{K)0},\\
\phit & = 0,\\
\wt^i & = \delta^i_J \wt^J,\\
\tilde{h}_{ij} &= -\delta^0_i\delta^0_j + \at\delta^I_i\delta^J_i \delta_{IJ},\\
\tilde{d}^{ijm}_{kl} &= \frac{\at'}{\at}\Bigl( -\frac{3}{2} \delta^{i}_k\delta^j_l \delta^m_0 +
\delta^{(i}_0 \delta^{j)}_k \delta^I_l\delta^m_I + \frac{1}{\at} \delta^{I(i} \delta^{j)}_k \delta^m_I\delta^0_l
-\delta^{(i}_I \delta^{j)}_k \delta^I_l \delta^m_0 \Bigr),\\
\omega^{ij}_0 & = \delta^i_0\delta^j_0\del_t\Phit,
\intertext{and}
\omega^{ij}_I & =  \Delta^{-1}\bigl(-\delta^i_0\delta^j_0 \delta_I\delta_J\Phit - \tilde{d}^{ijI}_{00}\del_I\Phit
 + 8\delta^{(i}_0\Pi(\rhot\wt^{j)}) + \tilde{h}_{kl}\ut^{kl}\Pi(\rhot)\bigr).
}
Then
\alin{limB1}{
\Wt &= (\ut^{ij}_0,\Wt_I^{ij},\ut^{ij},\phit,\alphat,\wt^i)^T, \\
\omega &= (\omega^{ij}_0,\omega^{ij}_I,0,0,0,0)^T,
}
defines a solution to the limit equations \eqref{limeqn1.1}-\eqref{limeqn1.2} on the spacetime region $M_0=[0,T_0)\times \Tbb^3$.
\end{prop}
\begin{proof}
The proof follows by a straightforward calculation that verifies $\Wt$ is a solution to
the limit equation \eqref{limeqn1.1}-\eqref{limeqn1.2}.
\end{proof}

%% file: nlim.tex
\sect{nlim}{The Newtonian limit}

We are now ready to prove Theorem \ref{nlimA}.

\begin{proof}[Proof of Theorem \ref{nlimA}]
Given
$\alphat_0 \in H^s$, $\wt^I_0\in H^s(\Rbb^3)$, $\uc^{IJ}_0 \in \Hb^{s}(\Sbb{3})$ ,
$\uc^{IJ} \in \Hb^{s+1}(\Sbb{3})$, we let
\alin{nlimA5}{
\alpha_\ep|_{t=0} & = \alphat_0, \\
(w^j_\ep)|_{t=0} &= (w^0_\ep,\wt^I_0), \\
(\ub^{ij}_\ep)\bigl|_{t=0} & = \begin{pmatrix} \uc^{00}_\ep & \frac{1}{\ep} y^J(\ep) + \ub^{0J}_\ep \\
\frac{1}{\ep}y^J(\ep) + \uc^{J0}_\ep & \ep \uc^{IJ} \end{pmatrix},\\
(\del_t\ub^{ij}_\ep)\bigl|_{t=0} &= \begin{pmatrix} \del_t\ub^{00}_\ep\bigl|_{t=0}  & \del_t\ub^{0J}_\ep|_{t=0} \\
\del_t\ub^{0J}_\ep\bigl|_{t=0} & \uc_0^{IJ} \end{pmatrix},
}
where
\alin{nlimA6}{
\del_t\ub^{00}_\ep\bigl|_{t=0} = -\frac{1}{\ep}\del_I \uc^{I0}_\ep - \frac{3}{2}\ln(a)' \uc^{00}_\ep -\frac{1}{2} a' \delta_{IJ} \ep \uc^{IJ},\\
\del_t\ub^{0J}_\ep\bigl|_{t=0} = -\del_I\uc^{IJ} - \frac{5}{2}\ln(a)'
\left(\frac{1}{\ep}y^{J}(\ep) + \uc^{0J}_\ep\right),
}
be the initial data from Theorem \ref{idataA}. By construction, this data solves the constraint equations
\eqref{ceqns.1}-\eqref{ceqns.3}, and depends analytically on $\ep$, and satisfies
\gath{nlimA7}{
\sigma_\ep  = \sigma_0 + \text{\rm O}(\ep), \quad
w^0_\ep = \text{\rm O}(\ep), \quad
y^J(\ep) =  y^J_0 + \text{\rm O}(\ep), \\
\uc^{00}_\ep =  \phi + \text{\rm O}(\ep),
\AND
\uc^{0J}_\ep  = \text{\rm O}(\ep),
}
where
\gath{nlimA8}{
\sigma_0 = \int_{\Tbb^3} \rho \, d^3 x,\quad
y^J_0  = - \frac{1}{\sigma_0}\int_{\Tbb^3}\rho w^J \, d^3 x,
\intertext{and}
\phi = \Delta^{-1}(\rho-\sigma_0).
}
A straightforward calculation then shows that
\leqn{nlimA9}{
W_\ep(t) = \bigl(u^{ij}_{0,\ep}(t), W^{ij}_{I,\ep}(t),u^{ij}_{\ep}(t),\phi_{\ep}(t),\alpha_\ep(t),w^j_{\ep}(t)\bigl)^T
}
satisfies
\leqn{nlimA10}{
W_0\bigl|_{t=0} = (\ut^{ij}_0|_{t=0} ,0,\ut^{ij}\bigl|_{t=0},0,\alphat_0,\delta^j_I \wt^I_0),
}
where
\alin{nlimA11}{
\rhot & = (4Kn(n+1))^{-n}\alphat^{2n}, \\
\at &= \exp\left( \int_{0}^{t}\left(\frac{8}{3} \int_{\Tbb^3} \rhot(s)  d^3 x\right)^{\frac{1}{2}} ds \right), \\
\mut & = \int_{\Tbb^3} \rhot \, d^3 x, \\
\zeta^J & = \int_{\Tbb^3} \rhot \wt^J \, d^3 x, \\
\ut^{ij}_0 & =2\delta_0^{(i}\delta_J^{j)} \left( -\left(\frac{1}{\mut}\zeta^J\right)' - \frac{\at'}{2 \at}\frac{1}{\mut}\zeta^J\right) \\
\ut^{ij} & = 2\delta_0^{(i}\delta_J^{j)}\left( -\frac{2}{\mut}\zeta^J\right),
}
and $\{\alphat,\wt^I,\Phit\}$ is the solution to the Cosmological Poisson-Euler equations from Proposition \ref{limA}
that is generated from the initial data $\alphat|_{t=0}=\alphat_0$ and $\wt^I|_{t=0}=\wt^I_0$.
Also, it can be verified using the evolution equations \eqref{Wdef} that our choice of initial data guarantees that
\leqn{nlimA12}{
\norm{\del_t W_\ep\bigl|_{t=0}}_{\Hc^s} \lesssim 1 \quad 0<\ep <\ep_0
}
where
\eqn{nlimA13}{
\Hc^s = \bigl(H^s(\Sbb{4})\bigr)^3\times \Rbb \times H^s\times H^s(\Rbb^4).
}
Together, Proposition \ref{limB} and \eqref{nlimA10}-\eqref{nlimA12} allow us to apply Theorem 1 of \cite{Scho88} (see also
the remarks in Section 1 of \cite{Scho88})
and conclude (shrinking $\ep_0$ if necessary) that for any $T<T_0$, there exists maps
\eqn{nlimA15}{
W_\ep \in \bigcap_{p=0}^{\ell+3}C^p\bigl([0,T),\Hc^{s-p}\bigr) \qquad 0<\ep < \ep_0, \\
}
such that
\begin{itemize}
\item[(i)] $W_\ep (t,x^I)$ solves equation \eqref{nonloc1} on the spacetime region $(t=x^0,x^I)\in D=[0,T)\times \Tbb^3$,
\item[(ii)] $W_\ep$ satisfy the estimates
\eqn{nlimA16}{
\norm{W_\ep(t)}_{\Hc^s} + \norm{\del_t W_\ep (t)}_{\Hc^{s-1}} \lesssim 1,
}
for all $(t,\ep) \in [0,T)\times (0,\ep_0)$, and
\item[(iv)]
\eqn{nlimA17}{
\norm{W_\ep(t)-\Wt(t)}_{H^{s-1}}  \lesssim \ep,
}
for all $(t,\ep) \in [0,T)\times (0,\ep_0)$, where $\Wt$ is the solution of the limit equations from
Proposition \ref{limB}.
\end{itemize}
Finally, we observe that similar arguments used in the proof of Proposition 6.1 in \cite{Oli06} show that
\eqn{nlimA18}{
\{\ub_\ep^{ij} = \ep^{-1} u^{ij}_\ep,\alpha_\ep,w^i_\ep\}
}
determines, via formulas  \eqref{dendef} and \eqref{metrecA}-\eqref{wdef.intro}, a solution to the
Einstein-Euler equations \eqref{EEeqn} in the harmonic gauge \eqref{harm1}, and moreover, that
\alin{nlimA19}{
 \ep \del_t \ub^{00}_\ep &=  u^{00}_{0,\ep}, \\
 \ep \del_t \ub^{0J} & =  u^{0J}_{0,\ep}  - \frac{a'}{2a} \ep \ub^{0J}_\ep, \\
\ep \del_t \ub^{IJ}  & =  u^{IJ}_{0,\ep} - \frac{a'}{a} \ep\ub^{IJ}_\ep,
}
and
\alin{nlimA20}{
\del_K \ub^{00}_\ep  &= W^{00}_{K,\ep} - a' \ep \ub^{0K}_\ep + \del_K\Phi_\ep, \\
 \del_K \ub^{0J}_\ep  & = W^{0J}_{K,\ep}  - \frac{a'}{2}\ep \ub^{KJ}_\ep+ \frac{a'}{2a} \delta^J_K \ep \ub^{00}_\ep, \\
 \del_K \ub^{IJ}_\ep & =  W^{IJ}_{K,\ep} - \frac{a'}{a} \ep \delta^{(I}_K \ub^{J)0}_\ep.
}
This combined with the statements (i)-(iii) above
completes the proof.
\end{proof}

%% file: cnl.bbl
\begin{thebibliography}{0}

\bibitem{BK07} U.~Brauer and L.~Karp, \emph{Local existence of classical solutions of the
system using weighted sobolev spaces of fractional order}, Les Comptes
l'Acad\'{e}mie des sciences / S\'{e}rie Mathematique {\bf 345} (2007), 49-54.
\bibitem{Deim} K.~Deimling, \emph{Nonlinear functional analysis}, Springer-Verlag, Berlin, 1998.
\bibitem{KM82} S.~Klainerman and A.~Majda, {\em Compressible and incompressible fluids},
Comm. Pure Appl. Math. {\bf 35} (1982), 629-651.
\bibitem{Kreiss80} H.O.~Kreiss, \emph{Problems with different time
scales for partial differential equations}, Comm. Pure Appl. Math.
{\bf 33} (1980), 399-439.
\bibitem{Kunz76} H.P.~K\"{u}nzle \emph{Covariant Newtonian limit of Lorentz space-times}, Gen. Rel. Grav.
{\bf 7} (1976), 445-457.
\bibitem{Lott} M.~Lottermoser, \emph{A convergent post-Newtonian approximation for the
constraints in general relativity}, Ann. Inst. Henri Poincar\'{e} \textbf{57} (1992), 279-317.
\bibitem{Mak} T.~Makino, ``On a local existence 
theorem for the evolution equation of gaseous stars'',
in {\em Patterns and Waves}, edited by T.~Nishida, 
M.~Mimura, and H.~Fujii, North-Holland, Amsterdam, 1986.
\bibitem{Oli06} T.A.~Oliynyk, {\em The Newtonian limit for perfect fluids}, Commun. Math. Phys.
{\bf 276} (2007), 131-188.
\bibitem{Oli07} T.A.~Oliynyk, \emph{Post-Newtonian expansions for perfect fluids},
Commun. Math. Phys. {\bf 288} (2009), 847-886.
\bibitem{Oli09} T.A.~Oliynyk, \emph{Cosmological post-Newtonian expansions to arbitrary order},
Commun. Math. Phys. (accepted) [preprint: arXiv:0908.2836].
\bibitem{Ren92} A.D.~Rendall, {\em The initial value problem
for a class of general relativistic fluid bodies},
J. Math. Phys. {\bf 33} (1992), 1047-1053.
\bibitem{RS97} C.~ R\"{u}ede and N.~Straumann, \emph{On Newton-Cartan cosmology}, Helv.Phys.Acta {\bf 70} (1997), 318-335. 
\bibitem{Scho86} S.~Schochet, {\em Symmetric hyperbolic systems with a large parameter},
Comm. partial differential equations, {\bf 11} (1986), 1627-1651.
\bibitem{Scho88}  S.~Schochet, {\em Asymptotics for symmetric hyperbolic systems with a large parameter}, 
J. differential equations {\bf 75} (1988), 1-27.
\bibitem{TayIII} M.E~Taylor, {\em Partial differential equations III, nonlinear equations}, Springer, New York, 1996.
\end{thebibliography}
